\documentclass[
    ,final            % use final for the camera ready runs
  ]
  {aipproc}

\layoutstyle{8x11single}

\begin{document}

\title{Relativistic Models for Quasi-Elastic Neutrino-Nucleus Scattering}

\classification{25.30.Pt, 13.15.+g, 24.10.Jv
{\bf [Presented by M.B. Barbaro at PANIC 2011, MIT, Cambridge, MA. July 2011]}
}
\keywords      {Neutrino interactions; Relativistic mean field theory;
Meson exhcange currents; Superscaling; Quasielastic scattering.}

\author{M.B. Barbaro}{
  address={
%Dipartimento di Fisica Teorica, 
Universit\`a di Torino and
  INFN, Sezione di Torino, 
%Via P. Giuria 1, 
10125 Torino, ITALY}
}

\author{J.E. Amaro}{
  address={
%Departamento de F\'{\i}sica At\'{o}mica, Molecular y Nuclear,
Universidad de Granada,
  18071 Granada, SPAIN}
}

\author{J.A. Caballero}{
  address={
%Departamento de F\'{\i}sica At\'{o}mica, Molecular y Nuclear,
Universidad de Sevilla, 41080 Sevilla, SPAIN} 
}

\author{T.W. Donnelly}{
  address={
%Center for Theoretical Physics, Laboratory for Nuclear Science 
CTP, LNS and Department of Physics, 
%Massachusetts Institute of Technology,
%CTP, LNS, 
MIT,  Cambridge, MA 02139, USA} 
}

\author{J.M. Ud\'{\i}as}{
  address={
%Grupo de F\'{\i}sica Nuclear, Departamento de  F\'{\i}sica At\'{o}mica, Molecular y Nuclear,
Universidad Complutense de Madrid,
  28040 Madrid, SPAIN}
}

\begin{abstract}
Two relativistic approaches to charged-current 
quasielastic neutrino-nucleus scattering are
illustrated and compared: one is phenomenological and based on the superscaling
behavior of electron scattering data and the other relies on the microscopic
description of nuclear dynamics in relativistic mean field theory. 
The role of meson exchange currents in the two-particle two-hole sector is 
explored. The predictions of the models for differential and total cross 
sections are presented and compared with the MiniBooNE data.
\end{abstract}

\maketitle

%%%%%%%%%%%%%%%%%%%%%%%%%%%%%%%%%%%%%%%%%%%%
%% MAINMATTER
%%%%%%%%%%%%%%%%%%%%%%%%%%%%%%%%%%%%%%%%%%%%

%\section{Introduction}

The recent MiniBooNE data on muon neutrino charged-current quasielastic (CCQE)
scattering~\cite{AguilarArevalo:2010zc} have raised an important
debate on the role played by both nuclear and nucleonic ingredients 
entering in the description of the reaction.
Unexpectedly, the cross section turns out to be substantially underestimated by
the Relativistic Fermi Gas (RFG) prediction, unless 
an unusually large {\em ad hoc} value of the axial mass $M_A\simeq$1.35 GeV/c$^2$ 
(as compared to the standard value $M_A\simeq$1 GeV/c$^2$) is employed
in the dipole parametrization of the nucleon axial form factor. 
From comparison with electron scattering data the RFG model is known, however, 
to be too crude to account for the nuclear dynamics: therefore this
result should be taken more as an indication of incompleteness of the 
theoretical description of the nuclear many-body problem rather 
than as a true indication for a larger axial mass. 

At the level of the impulse approximation (IA), a number of
much more sophisticated descriptions of the nuclear dynamics other
than the RFG also underpredict the measured CCQE cross section 
(see, e.g., Ref.~\cite{Cortona} for a full list of references, that we omit 
here for loss of space).
Possible explanations of this puzzle have been proposed in the literature,
based either on multinucleon knockout or on particular treatments of final 
state interactions through phenomenological optical potentials, indicating
that contributions beyond the simple IA play an important role in QE 
neutrino reactions.

Here we summarize the results of 
Refs.~\cite{Amaro:2004bs,Amaro:2010sd,Amaro:2011qb}, where the predictions of
the following two relativistic models were compared with each other and with 
the MiniBooNE data:
\begin{enumerate}
\item{the SuperScaling Approach (SuSA) including 2p2h Meson Exchange Currents (MEC);}
\item{the Relativistic Mean Field model (RMF).}
\end{enumerate}
Both models, although being far more realistic than the RFG, share with it 
the important property of treating exactly the relativistic
aspects of the problem: these cannot be neglected for the kinematics
of MiniBooNE, where the neutrino energy reaches values as high as 3
GeV.

The ``SuSA'' approach~\cite{Amaro:2004bs} is based
on the assumed universality of the scaling function for
electromagnetic and weak interactions. Analyses of inclusive $(e,e^\prime)$ 
data have demonstrated that at energy transfers below the 
QE peak superscaling is fulfilled rather well: this means
that the reduced cross section is largely independent of the
momentum transfer (I-kind scaling) and nuclear target
(II-kind scaling), when represented as a function of the
appropriate scaling variable. From these analyses a phenomenological
scaling function, dramatically different in size and
shape from the RFG parabola, has been extracted from the longitudinal 
QE electron scattering response and used to predict neutrino-nucleus cross 
sections by multiplying it by the corresponding elementary weak cross sections. 
The model reproduces
by construction the longitudinal electron scattering response at all kinematics
and for all nuclei. 
Its limitations come from the assumptions on which the approach is based, namely: 1) the equality of the longitudinal and transverse
scaling functions (0-kind scaling), a property violated by the L/T separated 
data, which show a transverse scaling function typically larger than the
longitudinal one;
2) the equality of the scaling functions in different isospin channels 
(III-kind scaling), which allows to use the longitudinal electron scattering
data (having both isoscalar and isovector components) to predict
the purely isovector CC neutrino cross section.

The results of the SuSA model for the double differential, single differential and total CCQE neutrino cross sections are shown in Figs.~1-3, 
where they appear to fall below the data for most of the angle and energy 
bins.
Note that we do not compare with the most forward angles
(0.9$<\cos\theta<$1) since for such kinematics roughly 1/2 of the
cross section has been proved~\cite{Amaro:2010sd} to arise from very low 
excitation energies (<50 MeV), where the cross section is dominated by 
collective excitations and any approach based on IA is bound to fail.

To go beyond the SuSA approach one must take into account
superscaling violations, which
occurr mainly in the transverse channel at energies above the QE peak 
and are associated to non-impulsive effects,
like inelastic scattering and meson-exchange currents.
The latter are two-body currents carried by a virtual meson
exchanged between two bound nucleons and can excite both 1p1h and 2p2h states.
In the 1p1h sector, studies of electromagnetic $(e,e^\prime)$ process have 
shown that the MEC, when combined with the corresponding correlations, which 
are needed 
to preserve gauge invariance, give a small contribution to the QE cross section
and can be neglected in first approximation.
On the other hand in the 2p2h sector the MEC are known to give a significant 
positive contribution to the $(e,e^\prime)$ cross section at high energy 
transfers, leading to a partial filling of the ``dip'' between the QE 
and $\Delta$-resonance peaks. 
This region is relevant for the MiniBooNE experiment, where
``QE'' events (namely with no real pions in the final state) 
can involve transferred energies far beyond the QE peak,
due to the large energy range spanned by the neutrino flux.

\begin{figure}
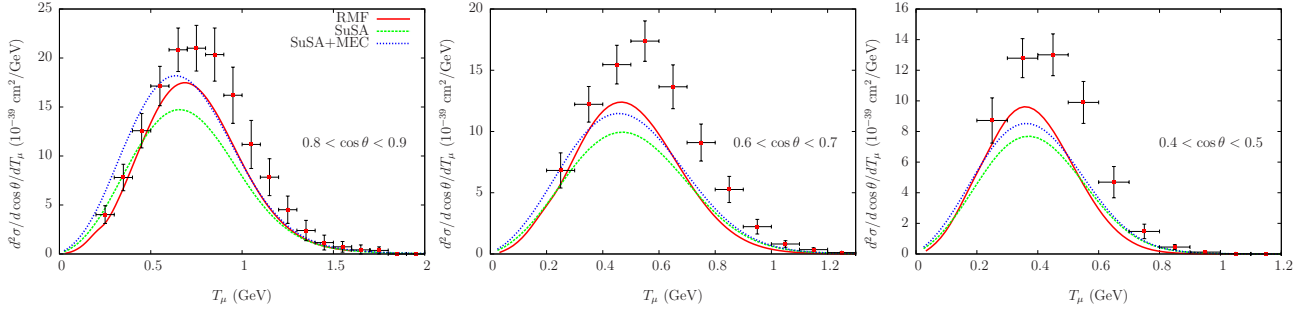

  \includegraphics[height=.18\textheight]{RMF_c085.epsi}
  \includegraphics[height=.18\textheight]{RMF_c065.epsi}
  \includegraphics[height=.18\textheight]{RMF_c045.epsi}
  \caption{ Flux-integrated $\nu_\mu$-$^{12}$C CCQE 
double differential cross section per target nucleon 
evaluated in the 
SuSA model with and without inclusion of 2p2h MEC and in the RMF
model and displayed
versus the muon kinetic energy $T_\mu$ for various bins of
the muon scattering angle $\cos\theta$.
Here and in the following figures the data are from MiniBooNE~\cite{AguilarArevalo:2010zc}.} 
\end{figure}
\begin{figure}
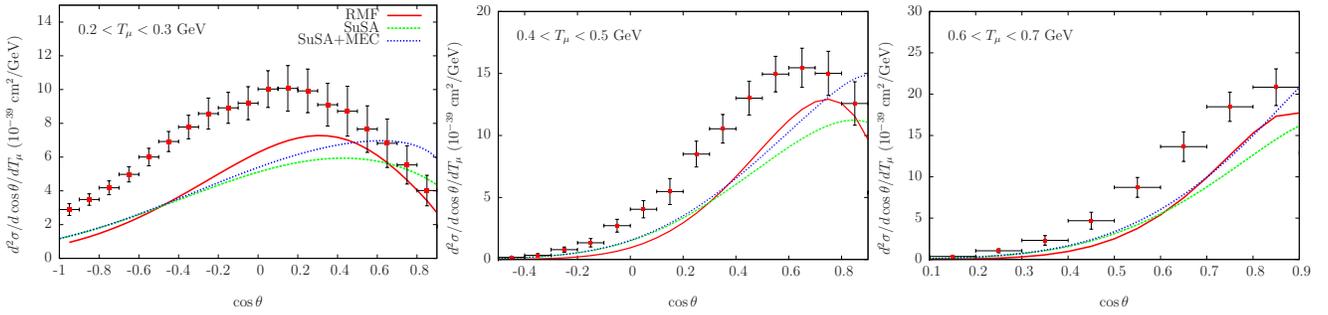

  \includegraphics[height=.18\textheight]{RMF_t025.epsi}
  \includegraphics[height=.18\textheight]{RMF_t045.epsi}
  \includegraphics[height=.18\textheight]{RMF_t065.epsi}
  \caption{Same as Fig.~1, but now displayed versus the scattering angle
$\cos\theta$ for various bins of $T_\mu$.
}
\end{figure}

In the results presented in Figs.~1-3 we have used a fully relativistic model,
developed for use in electron scattering studies, 
where all the MEC many-body diagrams 
containing two pionic lines that contribute to the electromagnetic 2p2h transverse response are taken into account. 
In order to apply the model to neutrino scattering,
we observe that in lowest order the 2p2h sector is not directly reachable 
for the axial-vector matrix elements. Hence at this order the MEC affect only
the transverse polar vector response. 
As shown in Figs.~1-3, the inclusion of 2p2h MEC
in the SuSA approach yields larger cross sections and accordingly better
agreement with the data, but theory still lies below the data
at larger angles where the cross sections are smaller. 
It should be noted, however, that 
the present approach still lacks the
contributions from the correlation diagrams associated with the MEC which
are required by gauge invariance; these might improve the agreement
with the data, as suggested by recent results for inclusive electron 
scattering~\cite{Amaro:2010iu}.

\begin{figure}
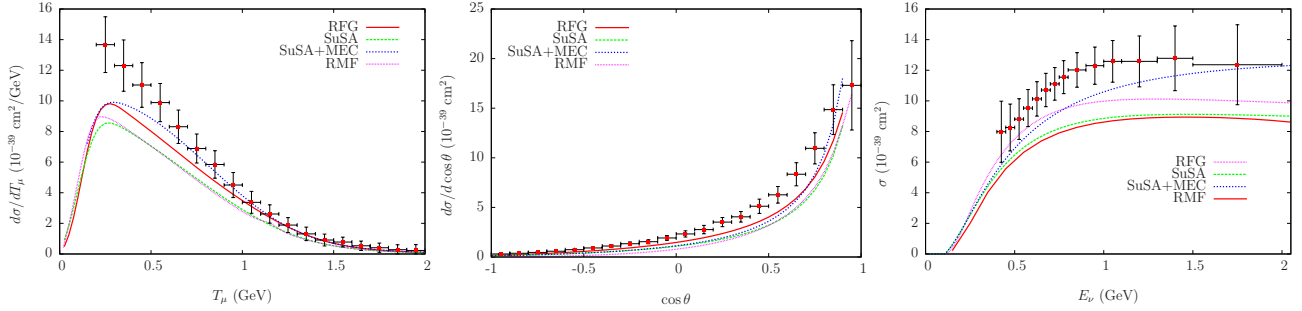

  \includegraphics[height=.18\textheight]{dsdt.epsi}
  \includegraphics[height=.18\textheight]{dsdcos.epsi}
  \includegraphics[height=.18\textheight]{Totalcs_Panic.epsi}
  \caption{Flux-averaged $\nu_\mu$-$^{12}C$ CCQE cross section 
integrated over the scattering angle and
displayed versus the muon kinetic energy (left panel),
integrated over the muon kinetic energy and 
displayed versus scattering angle 
(center panel), 
integrated over the muon kinetic energy and scattering angle and displayed
versus the unfolded neutrino energy 
(right panel). Beyond the models described in the text, the RFG result is also shown for comparison.}
\end{figure}

Before drawing definitive conclusions on the anomalous axial mass, it is
important to explore alternative approaches that have been shown to
be successful in describing inclusive QE $(e,e')$ processes.
This is the case for the RMF model, where a fully relativistic
description (kinematics and dynamics) of the process is
incorporated, and final state interactions are taken into account by using 
the same relativistic scalar and vector energy-independent potentials
considered in the description of the initial bound states. The RMF
model applied to inclusive QE $(e,e')$ processes has been shown to
describe the scaling behaviour and, in contrast with most other nuclear
models, to give rise to a
superscaling function with a significant asymmetry, in
complete accord with data.
Moreover, contrary to SuSA, where scaling of the zeroth kind is
assumed, the RMF model provides longitudinal and transverse scaling
functions which differ by typically $20\%$, the T one being larger.
When applied to the description of CCQE neutrino-nucleus cross sections,
the 0-kind
scaling violation introduced by the RMF approach, as well as the
different isospin character shown by the electromagnetic and weak
nucleon form factors, can lead to significant discrepancies between
the results provided by SuSA and RMF approaches. 
This is illustrated in Figs.~1, 2 and 3, where the differences between the
SuSA and RMF predictions are especially visible in the double differential
cross sections (Figs.~1 and 2), which are better described by the RMF model, 
and tend instead to be washed out by the integration (Fig.~3). 

Summarizing, we have applied two relativistic models, SuSA and RMF, both 
able to describe with good accuracy the longitudinal $(e,e^\prime)$ data,
to CCQE neutrino scattering, finding that both underestimate the MiniBooNE
cross sections: although the RMF does better than SuSA in reproducing the
shape of the double differential cross sections, the two approaches 
provide almost identical results for the single-differential and total
cross sections.
Although our scope here is not to extract a value for the axial mass of
the nucleon, but rather to understand which nuclear effects are 
effectively accounted for by a large axial cutoff parameter, let us
mention that a best fit of the RMF and SuSA results to the MiniBooNE
experimental cross section gives an effective axial mass
$M_A^{\rm eff}\simeq$ 1.5 GeV/c$^2$
and values in the range $1.35<M_A^{\rm eff}<1.65$ GeV/c$^2$ yield
results compatible with the MiniBooNE data within the experimental errors.

The inclusion of 2p2h MEC contributions in the SuSA approach
increases both the differential and the integrated cross sections and thus 
seems to improve the agreement with the data,
suggesting that the data can be explained without the need for a
large nucleon axial mass. However, in the present scheme, 
more refined calculations taking care of 
correlation currents and MEC effects in the axial-vector channel 
should be performed before final conclusions can be drawn.
We refer the reader to Refs.~\cite{Amaro:2004bs,Amaro:2010sd,Amaro:2011qb}
for further details and results.

%%%%%%%%%%%%%%%%%%%%%%%%%%%%%%%%%%%%%%%%%%%%%%%%
%% The bibliography can be prepared using the BibTeX program or
%% manually.
%%
%% The code below assumes that BibTeX is used.  If the bibliography is
%% produced without BibTeX comment out the following lines and see the
%% aipguide.pdf for further information.
%%
%% For your convenience a manually coded example is appended
%% after the \end{document}
%%%%%%%%%%%%%%%%%%%%%%%%%%%%%%%%%%%%%%%%%%%%%%%%

%%%%%%%%%%%%%%%%%%%%%%%%%%%%%%%%%%%%%%%%%%%%%%%%
%% You may have to change the BibTeX style below, depending on your
%% setup or preferences.
%%
%%
%% For The AIP proceedings layouts use either
%%%%%%%%%%%%%%%%%%%%%%%%%%%%%%%%%%%%%%%%%%%%

\bibliographystyle{aipproc}   % if natbib is available
%\bibliographystyle{aipprocl} % if natbib is missing

%%%%%%%%%%%%%%%%%%%%%%%%%%%%%%%%%%%%%%%%%%%
%% You probably want to use your own bibtex database here
%%%%%%%%%%%%%%%%%%%%%%%%%%%%%%%%%%%%%%%%%%%
%\bibliography{sample}

\begin{thebibliography}{9}

\bibitem{AguilarArevalo:2010zc}
  A.~A. Aguilar-Arevalo {\it et al.}  [MiniBooNE Collaboration],
  %``First Measurement of the Muon Neutrino Charged Current Quasielastic Double
  %Differential Cross Section,''
  \emph{Phys. Rev.}  \textbf{D81} 092005 (2010).
  %arXiv:1002.2680 [hep-ex].
  %%CITATION = ARXIV:1002.2680;%%
%
%\cite{Barbaro:2011gs}
\bibitem{Cortona}
  M.~B. Barbaro,
  %``Nuclear effects in charged-current quasielastic neutrino-nucleus
  %scattering,''
  Proceedings of "XIII Convegno di Cortona su Problemi di Fisica Nucleare 
  Teorica", 
%Cortona (Italy), April 6-8, 2011, 
  arXiv:1108.2732 [nucl-th].
  %%CITATION = ARXIV:1108.2732;%%
%
\bibitem{Amaro:2004bs}
  J.~E. Amaro, M.~B. Barbaro, J.~A. Caballero, T.~W. Donnelly, A. Molinari and I. Sick,
  %``Using electron scattering superscaling to predict charge-changing  neutrino
  %cross sections in nuclei,''
  \emph{Phys. Rev.}  \textbf{C71} 015501 (2005).
  %%CITATION = PHRVA,C71,015501;%%

\bibitem{Amaro:2010sd}
  J.~E. Amaro, M.~B. Barbaro, J.~A. Caballero, T.~W. Donnelly and C.~F. Williamson,
  %``Meson-exchange currents and quasielastic neutrino cross sections in the
  %SuperScaling Approximation model,''
  \emph{Phys. Lett.} \textbf{B696} 151 (2011). 
%  [arXiv:1010.1708 [nucl-th]].
  %%CITATION = PHLTA,B696,151;%%
% 44
\bibitem{Amaro:2011qb}
  J.~E. Amaro, M.~B. Barbaro, J.~A. Caballero, T.~W. Donnelly and 
  J.~M. Ud\'{\i}as,
  %``Relativistic analyses of quasielastic neutrino cross sections at MiniBooNE
  %kinematics,''
  \emph{Phys. Rev.}  {\bf D84} 033004 (2011).
%  arXiv:1104.5446 [nucl-th].
  %%CITATION = ARXIV:1104.5446;%%
%\cite{Amaro:2010iu}
\bibitem{Amaro:2010iu}
  J.~E.~Amaro, C.~Maieron, M.~B.~Barbaro, J.~A.~Caballero and T.~W.~Donnelly,
  %``Pionic correlations and meson-exchange currents in two-particle emission
  %induced by electron scattering,''
  \emph{Phys. Rev.}  \textbf{C82} 044601 (2010).
%  [arXiv:1008.0753 [nucl-th]].
  %%CITATION = PHRVA,C82,044601;%%

\end{thebibliography}

%%%%%%%%%%%%%%%%%%%%%%%%%%%%%%%%%%%%%%%%%%%
%% Just a reminder that you may have to run bibtex
%% All of it up to \end{document} can be removed
%% if you don't like the warning.
%%%%%%%%%%%%%%%%%%%%%%%%%%%%%%%%%%%%%%%%%%%
%
%\IfFileExists{\jobname.bbl}{}
% {\typeout{}
%  \typeout{******************************************}
%  \typeout{** Please run "bibtex \jobname" to optain}
%  \typeout{** the bibliography and then re-run LaTeX}
%  \typeout{** twice to fix the references!}
%  \typeout{******************************************}
%  \typeout{}
% }

%%%%%%%%%%%%%%%%%%%%%%%%%%%%%%%%%%%%%%%%%%%
%% The following lines show an example how to produce a bibliography
%% without the help of the BibTeX program. This could be used instead
%% of the above.
%%%%%%%%%%%%%%%%%%%%%%%%%%%%%%%%%%%%%%%%%%%

\end{document}